\documentclass[prd,preprintnumbers,floatfix,aps,nofootinbib,notitlepage]{revtex4}

\usepackage{latexsym}
\usepackage{epsfig}
\usepackage{amssymb}
\usepackage{amsmath}
\usepackage{dcolumn}
\usepackage{bm}

\newcommand{\lp}{\left(}
\newcommand{\rp}{\right)}

\newcommand{\ba}{\begin{eqnarray}}
\newcommand{\ea}{\end{eqnarray}}
\newcommand{\be}{\begin{equation}}
\newcommand{\ee}{\end{equation}}

\newcommand{\F}{f^{(0)}_{,RR}}
\newcommand{\f}{f^{(0)}_{,R\hat R}}
\newcommand{\Q}{f^{(0)}_{,\hat Q}}

\begin{document}

\title{Ghosts in pure and hybrid formalisms of gravity theories: a unified analysis}

\author{Tomi S. Koivisto}\email{tomi.koivisto@fys.uio.no}
\affiliation{Institute of Theoretical Astrophysics, University of Oslo, P.O. Box 1029 Blindern, N-0315 Oslo, Norway}
\author{Nicola Tamanini}\email{n.tamanini.11@ucl.ac.uk}
\affiliation {Department of Mathematics, University College London, Gower Street, London, WC1E 6BT, UK}

\date{\today}

\begin{abstract}

In the first order formalism of gravitational theories, the spacetime connection is considered as an independent variable to vary together with the metric. However,
the metric still generates its Levi-Civita connection that turns out to determine the geodesics of matter. Recently, ''hybrid'' gravity theories have been introduced
by constructing actions involving both the independent Palatini connection and the metric Levi-Civita connection. In this study a method is developed to analyse the 
field content of such theories, in particular to determine whether the propagating degrees of freedom are ghosts or tachyons. New types of second, fourth and
sixth order derivative gravity theories are investigated and the so called $f(X)$ theories are singled out as a viable class of ''hybrid'' extensions of General Relativity.

\end{abstract}

\maketitle

\section{Introduction}

Motivated by the oddities discovered by recent astronomical observations, a plethora of modified theories of gravity has been advanced during the last years. Their main scope is to reproduce the observed behavior of our universe without invoking any undetected entity such as dark matter, dark energy or the inflaton field. Although some of them may succeed in reaching the goal, some others give rise to unphysical features such as, for example, the appearance of ghosts.

Among all these modified theories, the most considered are the so-called $f(R)$ theories of gravity, where the action is taken to depend on a general function of the Ricci scalar $R$; see \cite{SF10} for some reviews. Their popularity comes from the fact that they are both theoretically viable and sufficiently simple to study in cosmology and other frameworks. To these ends, the $f(R)$ action is usually varied with respect to the metric tensor in order to produce the field equations governing the dynamics of gravitation. This approach is known under the name of metric (sometimes second-order) variational principle and it is the one first employed by Hilbert to derive the Einstein field equations. However a second variational method is well known since Einstein formulated it, though for historical reasons \cite{FFR82} it has been named Palatini (sometimes first-order) variational principle. It consists of an independent variation with respect of the metric and the (torsionless) connection. If applied to the Einstein-Hilbert action it gives back the Einstein field equations, meaning there are no physical difference for general relativity. However when considered for other theories, such as $f(R)$ gravity, it produces completely new dynamical field equations resulting in a different phenomenology; see \cite{SF10,Olmo:2011uz} for reviews.

The metric and Palatini approaches have been recently combined\footnote{See Refs.\cite{Amendola:2010bk,unified} for other frameworks for unifying the variational principles.} to give born to a new class of modified gravitational theories which has been named hybrid metric-Palatini or $f(X)$ gravity \cite{fxliterature}. The action is taken to depend linearly on the metric curvature scalar $R$ but nonlinearly on the Palatini curvature scalar $\hat R$ which is modulated by an arbitrary function in analogy with Palatini $f(R)$ theories. Some cosmological and astrophysical applications of these theories have already been studied and it has been shown that they can both pass the solar system experiments and predict a late time accelerated cosmological expansion. For this reason they can constitute a promising theoretical explanation for dark energy. The hybrid metric-Palatini approach has then been generalized considering a general function of both $R$ and $\hat R$ \cite{Tamanini:2013ltp}, independently from an earlier work which considered similar ideas for different reasons \cite{Flanagan:2003iw}. The cosmological evolution of these generalized theories have also been analyzed and late time acceleration can be achieved in this context. Moreover it has been proved that this class of theories is dynamical equivalent to a nonminimally coupled biscalar theory of gravity and that one of the two scalar disappears in the $f(X)$ subclass. In other words, the theory introduces two new scalar degrees of freedom, one in the $f(X)$ case, on top of the metric ones.

In this paper we will consider a further generalization of gravitational theories within the hybrid metric-Palatini framework. In particular we will take the gravitational action to arbitrarily depend on $R$, $\hat R$ and $\hat Q_H = R^{\mu\nu}\hat{R}_{\mu\nu}$. This generalization is done in analogy with the Ricci square modified gravity which has been largely studied in both metric \cite{Stelle:1976gc,metricRicciSquare} and Palatini \cite{PalatiniRicciSquare} formulations. Note however that the invariants $\hat Q_H$ we are considering can only appear within a hybrid metric-Palatini context, and thus were never considered before. Our analysis aims at characterizing the flat space propagator in order to verify the presence of ghosts for such theories. It is a well known result that Ricci square modifications always leads to a spin-2 ghost called the Weyl ghost \cite{Stelle:1976gc}. It is thus interesting to perform the ghost analysis also for this class of hybrid theories in order to check if also these models are inevitably haunted by ghosts.

The paper is organized as follows. In Sec.~\ref{sec:act&eqs} we will introduce the action of the theory and derive the field equations discussing how they can be solved in general. In Sec.~\ref{sec:weakfield} we will consider an expansion around Minkowski space and compute the field equations in this weak field limit. In Sec.~\ref{sec:propagators} we will then invert the field equations to calculate the flat space propagator. An analysis for several subcases included in the class of theories we consider will be briefly carried out in order to check that the results are consistent with previous works, and a health diagnosis will be performed for the new theories included in our action. Finally we draw our conclusions in Sec.~\ref{sec:concl}. In what follows we will set $\kappa^2 = 8\pi G/c^4$ for the sake of simplicity.

%HYBRID THEORIES... \cite{fxliterature}\\
%RICCI SQUARE IN PALATINI...

\section{Action and Field Equations}
\label{sec:act&eqs}

The action we will adopt for our analysis is given by
\begin{align}
S = \frac{1}{2\kappa^2} \int d^4x \sqrt{-g}\, f( R,\hat R,\hat Q_H ) \,,
\label{001}
\end{align}
where $R=g^{\mu\nu}R_{\mu\nu}$ and $\hat R=g^{\mu\nu}\hat R_{\mu\nu}$ are the metric and Palatini curvature scalars respectively. The first one is composed with the usual Levi-Civita connection $\Gamma_{\mu\nu}^\lambda$ and it is completely determined by the metric $g_{\mu\nu}$,
\begin{equation}
{R}_{\mu\nu} \equiv {\Gamma}^\alpha_{\mu\nu ,\alpha} -
{\Gamma}^\alpha_{\mu\alpha , \nu} +
{\Gamma}^\alpha_{\alpha\lambda}{\Gamma}^\lambda_{\mu\nu}
-{\Gamma}^\alpha_{\mu\lambda}{\Gamma}^\lambda_{\alpha\nu}\,,
\end{equation}
 the second one depends on an independent torsionless connection $\hat\Gamma^\lambda_{\mu\nu}$ forming the following Ricci tensor
\begin{equation}
\hat{R}_{\mu\nu} \equiv \hat{\Gamma}^\alpha_{\mu\nu ,\alpha} -
\hat{\Gamma}^\alpha_{\mu\alpha , \nu} +
\hat{\Gamma}^\alpha_{\alpha\lambda}\hat{\Gamma}^\lambda_{\mu\nu}
-\hat{\Gamma}^\alpha_{\mu\lambda}\hat{\Gamma}^\lambda_{\alpha\nu}\,.
\end{equation}
The function $f$ in action (\ref{001}) arbitrarily depends on $R$, $\hat R$ and
\begin{align}
\hat Q_H = R^{\mu\nu}\hat{R}_{\mu\nu} \,,
\end{align}
which will be denoted as the {\it hybrid Ricci square invariant}. Note that, though $\hat R_{\mu\nu}$ is in general asymmetric, only its symmetric part enters the action (\ref{001}).

Action (\ref{001}) generalizes the theory considered in \cite{Tamanini:2013ltp} where the hybrid Ricci square invariant was absent. As shown in \cite{Tamanini:2013ltp,Flanagan:2003iw} such a theory %where no higher order curvature invariants appear 
is dynamically equivalent to a nonminimally coupled biscalar theory. 
% which, having no higher order derivatives (and assuming a good behaviour of the scalar fields), turns out to be ghosts free (IS THIS TRUE???).
Here we undertake the study of viability of such a biscalar theory. It is interesting to consider also theories with other higher order invariants in the hybrid metric-Palatini context. Although several new invariants can be built within this framework
%\footnote{In the hybrid metric-Palatini approach there are up to 9 possible Ricci square invariants; 12 if also pseudoinvariants are considered. This is due to the fact that for the Palatini connection there are two possible second-order (Ricci) asymmetric tensors which can be formed with contractions of the Riemann tensor and the metric.}
, we decide to work only with $\hat Q_H$ since it is the simpler among the higher curvature invariants not appearing in pure metric and Palatini gravities and thus never considered in the previous literature.
%\footnote{The other one is $R_{\mu\nu}\hat B^{\mu\nu}$ where $\hat B_{\mu\nu} =g^{\alpha\beta}\hat R_{\alpha\mu\beta\nu}$.}. 
%Moreover the theory described by action (\ref{001}) is enough simple to make the ghosts analysis possible without specifiying a matter action as it happens in Palatini (symmetric) Ricci square gravity. 

Variation of (\ref{001}) with respect to the metric produces the following field equations
\begin{multline}
f_{,R} R_{\mu\nu} -\frac{1}{2}g_{\mu\nu}f +g_{\mu\nu}\Box_g f_{,R} -\nabla_\mu\nabla_\nu f_{,R} +f_{,\hat{R}} \hat{R}_{\mu\nu} \\
+2f_{,\hat{Q}}R_\mu^\lambda \hat{R}_{\nu\lambda} +\frac{1}{2}\Box_g\left(f_{,\hat{Q}}\hat{R}_{\mu\nu}\right) +\frac{1}{2}g_{\mu\nu}\nabla_\alpha\nabla_\beta\left(f_{,\hat{Q}} \hat{R}^{\alpha\beta}\right) -\nabla_\lambda\nabla_{(\nu}\left(f_{,\hat{Q}} \hat{R}_{\mu)}^{\lambda}\right) = \kappa^2 T_{\mu\nu} \,,
\label{003}
\end{multline}
where $f_{,R}$, $f_{,\hat{R}}$ and $f_{,\hat{Q}}$ are the derivatives of $f$ with respect to $R$, $\hat R$ and $\hat Q_H$ respectively. Round brackets between two indices denotes symmetrization and $\Box_g=g^{\mu\nu}\nabla_\mu\nabla_\nu$.
The energy-momentum tensor on the right hand side of (\ref{003}) has been derived adding a standard matter action to (\ref{001}).
Variation of action (\ref{001}) with respect to the independent connection $\hat\Gamma_{\mu\nu}^\lambda$ yields, after some manipulations, the condition
\begin{align}
\hat\nabla_\lambda\left(\sqrt{-g}f_{,\hat{R}}g^{\mu\nu} +\sqrt{-g}f_{,\hat{Q}}R^{\mu\nu}\right) =0 \,,
\label{005}
\end{align}
where $\hat\nabla_\mu$ denotes the covariant derivative of $\hat\Gamma_{\mu\nu}^\lambda$. In order to solve this equation a new metric $\hat g_{\mu\nu}$ implicitly defined by
\begin{align}
\sqrt{-\hat g}\,\hat g^{\mu\nu} = \sqrt{-g}\left(f_{,\hat{R}}g^{\mu\nu} +f_{,\hat{Q}}R^{\mu\nu}\right) \,,
\label{002}
\end{align}
emerges in Palatini theories. 
%The relation between the metrics is conformal if the theory involves only the invariant $\hat{R}$; the disformal relation is generic in gravity and scalar field theories as argued in \cite{Koivisto:2012za} (??).  
If the right hand side of (\ref{002}) is independent of the connection $\hat\Gamma_{\mu\nu}^\lambda$ a solution is given by
\begin{align}
\hat\Gamma^\lambda_{\mu\nu} = \frac{1}{2}\hat{g}^{\lambda\sigma} \left(\partial_\mu\hat{g}_{\sigma\nu} +\partial_\nu\hat{g}_{\sigma\mu} -\partial_\sigma\hat{g}_{\mu\nu}\right) \,,
\label{004}
\end{align}
or, in other words, the independent connection is given by the Levi-Civita connection in terms of the metric $\hat{g}_{\mu\nu}$. However in (\ref{002}) the two functions $f_{\hat{R}}$ and $f_{\hat{Q}}$ still implicitly depends on $\hat\Gamma_{\mu\nu}^\lambda$. Within the purely Palatini framework one usually employs field equations (\ref{003}) in order to algebraically relate $\hat R$ and $\hat Q$ to the matter fields \cite{PalatiniRicciSquare}. In this way one can substitute these invariants with matter fields wherever they appear in the field equations. In particular replacing them inside the condition (\ref{002}) allows to solve for the independent connection as in (\ref{004}). In our case however this cannot be done in general since in the field equations (\ref{003}) derivatives of $\hat Q_H$ explicitly appear and thus an algebraic relation with matter fields is impossible to achieve. Although this prevents finding general solutions of the field equations in a simple algebraic way, we will see in the next section that in the weak field limit needed for the ghosts analysis the right hand side of (\ref{002}) will be independent of $\hat\Gamma_{\mu\nu}^\lambda$ at first order in the perturbations.

The explicit form of the (inverse) metric $\hat g^{\mu\nu}$ can be read off from Eq.(\ref{002}) by taking its determinant. Defining
\begin{align}
r^{\mu\nu} = f_{,\hat{R}}g^{\mu\nu} +f_{,\hat{Q}}R^{\mu\nu} \,,
\end{align}
and $1/r=\det(r^{\mu\nu})$ we find
\begin{align}
\hat g^{\mu\nu} = \frac{\sqrt{-r}}{\sqrt{-g}}\,r^{\mu\nu} \,.
\label{007}
\end{align}

\section{Weak Field Limit}
\label{sec:weakfield}

We now consider an expansion around Minkowski spacetime. The metric tensor will be then given by
\begin{align}
g_{\mu\nu} = \eta_{\mu\nu} +h_{\mu\nu} \,, \quad\mbox{with}\quad |h_{\mu\nu}|\ll 1 \,,
\end{align}
where $\eta_{\mu\nu}$ is the Minkowski metric. All the standard properties of the weak field limit approximations will hold, in particular $h_{\mu\nu}$ is Lorentz covariant and can be treated as a Lorentz tensor on Minkowski spacetime. Up to second order in $h_{\mu\nu}$ we have $g^{\mu\nu}=\eta^{\mu\nu}-h^{\mu\nu}+o(h^2)$ and
\begin{align}
g = -1 +h +o(h^2) \,,
\end{align}
where we define $h=\eta^{\mu\nu}h_{\mu\nu}$ and denote with $o(h^2)$ second and higher orders in $h_{\mu\nu}$.

At this point we need to expand $f$ and its derivatives. Since these are functions of $R$, $\hat R$ and $\hat Q_H$ we need first to know which order in $h_{\mu\nu}$ are these invariants. The metric curvature scalar $R$ is of course of order one and has no zeroth order term as it happens in GR.
The Ricci tensor formed with the Palatini connection do not depend explicitly on the metric. However if one had a solution of $\hat\Gamma_{\mu\nu}^\lambda$ in terms of $g_{\mu\nu}$ this problem would not be present. In what follows we will assume that also $\hat R^{\mu\nu}$ is of order one in $h_{\mu\nu}$ and that its zeroth order term vanishes. This ansatz will be verified at the end of the calculations where we will indeed show that this is true. With this assumption we have $\hat R \sim o(h)$ and $\hat Q_H \sim o(h^2)$. The expansion of $f$ (and thus of all its derivatives) will be given by
\begin{align}
f(R,\hat R,\hat Q_H) = f^{(0)} +f^{(0)}_{,R}R^{(1)} +f^{(0)}_{,\hat R}\hat{R}^{(1)} +o(h^2) \,,
\end{align}
where $f^{(0)}$, $f^{(0)}_{,R}$ and $f^{(0)}_{,\hat R}$ are the constant values of the functions evaluated at $R=\hat R=\hat Q_H=0$ and $R^{(1)}$, $\hat R^{(1)}$ are the first order terms of $R$ and $\hat R$ respectively.

In what comes next we will make another ansatz, namely $f^{(0)}_{,\hat R\hat R}=0$ or, in other words, the second derivative of $f$ with respect to $\hat R$ has to vanish when evaluated at $R=\hat R=\hat Q_H=0$. This assumption is needed in order to solve (\ref{005}) in the weak field limit approximation, i.e.~it allows $r^{\mu\nu}$ to be $\hat\Gamma_{\mu\nu}^\lambda$ independent in this limit. Note that this puts a constraint on the general expression $f$ can assume. If $f$ is independent or linear in $\hat R$ this assumption is automatically satisfied. However this is true for more general functions $f$ and thus we will still consider a general $\hat R$ dependency.

Within these assumptions the expansion of $r^{\mu\nu}$ is given by
\begin{align}
r^{\mu\nu} = f^{(0)}_{,\hat R}\eta^{\mu\nu} -f^{(0)}_{,\hat R}h^{\mu\nu} +f^{(0)}_{,\hat RR}R^{(1)}\eta^{\mu\nu} +f^{(0)}_{,\hat Q} R^{(1)}{}^{\mu\nu} +o(h^2) \,.
\label{006}
\end{align}
One can immediately notice that now $r^{\mu\nu}$ is $\hat\Gamma_{\mu\nu}^\lambda$ indepedent up to first order in $h_{\mu\nu}$. Thanks to (\ref{006}) we can also find the expansion of $1/r$ which is given by
\begin{align}
\frac{1}{r} = {f^{(0)}_{,\hat R}}^4 \left(-1-h+ \frac{4f^{(0)}_{,\hat{R}R}+f^{(0)}_{,\hat Q}}{f^{(0)}_{,\hat R}}R^{(1)}\right) +o(h^2) \,.
\end{align}
We are now ready to expand $\hat g^{\mu\nu}$ given by (\ref{007}). Up to first order in $h_{\mu\nu}$ we find
\begin{align}
\hat g^{\mu\nu} = {f^{(0)}_{,\hat R}}^3 \left(\eta^{\mu\nu}-\hat h^{\mu\nu}\right) +o(h^2) \,,
\end{align}
where
\begin{align}
\hat h^{\mu\nu} = h^{\mu\nu} +\frac{6f^{(0)}_{\hat RR} +f^{(0)}_{,\hat Q}}{2f^{(0)}_{,\hat R}}R^{(1)}\eta^{\mu\nu} +\frac{f^{(0)}_{,\hat Q}}{f^{(0)}_{\hat R}}R^{(1)}{}^{\mu\nu} \,.
\end{align}
The inverse of $\hat g^{\mu\nu}$ is given by
\begin{align}
\hat g_{\mu\nu}= {f^{(0)}_{,\hat R}}^{-3} \left(\eta_{\mu\nu}+\hat h_{\mu\nu}\right) +o(h^2) \,,
\label{008}
\end{align}
and one can indeed verify that $\hat g^{\mu\alpha}\hat g_{\alpha\nu}=\delta^\mu_\nu +o(h^2)$. We can now give the solution of $\hat\Gamma_{\mu\nu}^\lambda$ up to first order in $h_{\mu\nu}$ using (\ref{004}). The result is
\begin{align}
\hat\Gamma^\lambda_{\mu\nu} = \frac{1}{2}\eta^{\lambda\sigma} \left(\partial_\mu\hat h_{\nu\sigma} +\partial_\nu\hat h_{\mu\sigma} -\partial_\sigma\hat h_{\mu\nu}\right) +o(h^2) \,.
\label{009}
\end{align}
Note that the constant factor in front of (\ref{008}) is completely unimportant since cancels out in $\hat\Gamma_{\mu\nu}^\lambda$ and then will not appear in the field equations. This is in agreement with the fact that $\hat g_{\mu\nu}$ can always be rescaled by a constant factor without altering the physics.

The expression (\ref{009}) for $\hat\Gamma_{\mu\nu}^\lambda$ allows us to write down the expansion of $\hat R_{\mu\nu}$ as
\begin{align}
\hat R_{\mu\nu} =\hat R^{(1)}_{\mu\nu} +o(h^2) = \partial_\sigma\partial_{(\nu}\hat h_{\mu)}^\sigma -\frac{1}{2}\partial_\mu\partial_\nu\hat h -\frac{1}{2}\Box\hat h_{\mu\nu} +o(h^2) \,,
\end{align}
where of course $\hat h=\eta^{\mu\nu}\hat h_{\mu\nu}$. One can now realize that our previous assumptions that $\hat R^{\mu\nu}$ has a vanishing zeroth order term is indeed verified. We can now expand $R^{(1)}_{\mu\nu}$ in terms of $h_{\mu\nu}$. Defining
\begin{align}
A = \frac{6f^{(0)}_{\hat RR} +f^{(0)}_{,\hat Q}}{2f^{(0)}_{,\hat R}} \,, \quad\mbox{and}\quad B=\frac{f^{(0)}_{,\hat Q}}{f^{(0)}_{\hat R}} \,,
\end{align}
we find
\begin{align}
\hat R^{(1)}_{\mu\nu} = R^{(1)}_{\mu\nu} -A\partial_\mu\partial_\nu\partial_\alpha\partial_\beta h^{\alpha\beta} -\frac{A}{2}\eta_{\mu\nu}\Box\partial_\alpha\partial_\beta h^{\alpha\beta} +\left(A+\frac{B}{4}\right)\Box\partial_\mu\partial_\nu h +\frac{A}{2}\eta_{\mu\nu}\Box^2h -\frac{B}{2}\partial_\alpha\partial_{(\mu}h_{\nu)}^\alpha +\frac{B}{4}\Box^2h_{\mu\nu} \,.
\end{align}
We have now all the ingredients to expand the field equations (\ref{003}) up to first order in $h_{\mu\nu}$. Following \cite{Biswas:2011ar,Biswas:2013ds} we can write them in the form
\begin{multline}
\frac{f^{(0)}}{2}\left(\eta_{\mu\nu}+h_{\mu\nu}\right) +a(\Box)\Box h_{\mu\nu} +2b(\Box) \partial_\sigma\partial_{(\mu}h_{\nu)}^\sigma +c(\Box) \left( \eta_{\mu\nu} \partial_\alpha\partial_\beta h^{\alpha\beta} + \partial_\mu\partial_\nu h \right) \\
+d(\Box)\eta_{\mu\nu}\Box h +\frac{e(\Box)}{\Box} \partial_\mu\partial_\nu\partial_\alpha\partial_\beta h^{\alpha\beta} = -2\kappa^2\,T_{\mu\nu} \,,
\label{010}
\end{multline}
where $a(\Box)$, $b(\Box)$, $c(\Box)$, $d(\Box)$ and $e(\Box)$ are functions of $\Box=\eta^{\mu\nu}\partial_\mu\partial_\nu$. In these equations $f^{(0)}$ represents the cosmological constant and will not influence the upcoming analysis. In our case the functions appearing in (\ref{010}) are given by
\begin{align}
a(\Box) &= f^{(0)}_{,R}+f^{(0)}_{,\hat R} -f^{(0)}_{,\hat Q}\frac{B}{4}\Box^2 \,, \label{abox} \\
b(\Box) &= -f^{(0)}_{,R}-f^{(0)}_{,\hat R} +f^{(0)}_{,\hat Q}\frac{B}{4}\Box^2 \,, \\
c(\Box) &= f^{(0)}_{,R}+f^{(0)}_{,\hat R} -2\left(f^{(0)}_{,RR}+4f^{(0)}_{,R\hat R}+f^{(0)}_{,\hat Q}\right)\Box +\left[f^{(0)}_{,R\hat R}\left(6A+B\right) +f^{(0)}_{,\hat Q}\left(2A+\frac{B}{4}\right)\right]\Box^2 \,, \\
%c_2(\Box) &= \\
d(\Box) &= -f^{(0)}_{,R}-f^{(0)}_{,\hat R} +2\left(f^{(0)}_{,RR}+4f^{(0)}_{,R\hat R}+f^{(0)}_{,\hat Q}\right)\Box -\left[f^{(0)}_{,R\hat R}\left(6A+B\right) +f^{(0)}_{,\hat Q}\left(2A+\frac{B}{4}\right)\right]\Box^2 \,, \\
e(\Box) &= 2\left(f^{(0)}_{,RR}+4f^{(0)}_{,R\hat R}+f^{(0)}_{,\hat Q}\right)\Box -\left[f^{(0)}_{,R\hat R}\left(6A+B\right) +f^{(0)}_{,\hat Q}\left(2A+\frac{B}{2}\right)\right]\Box^2 \,.
\end{align}
%(IS THE $f_Q$ TERM LINEAR IN $\Box$ ACCEPTABLE??)\\
We can immediately notice that the following relations hold
\begin{align}
a+b=0 \,, \quad c+d=0 \,, \quad b+c+e=0 \,.
\end{align}
At first order these constraints must be true in any gravitational theory preserving diffeomorphism invariance inasmuch as they directly follow from the Bianchi identities \cite{Biswas:2011ar,Biswas:2013ds}. Note also that in the GR limit $f^{(0)}_{,R}\rightarrow 1$, $f^{(0)}_{,\hat R}\rightarrow 0$, $f^{(0)}_{,\hat Q}\rightarrow 0$ (or equivalently $f^{(0)}_{,R}\rightarrow 0$, $f^{(0)}_{,\hat R}\rightarrow 1$, $f^{(0)}_{,\hat Q}\rightarrow 0$) we recover the right values $a=-b=c=-d=1$ and $e=0$.

With the field equations expanded at first order in $h_{\mu\nu}$ we can now compute the flat space propagator and check if ghosts appear in the theory.

\section{Flat Space Propagators}
\label{sec:propagators}

In this section, we will analyze the field content and viability of the theories contained in our general class of actions. We adopt the formalism presented in Ref. \cite{Biswas:2013ds}, that was originally introduced by van Niewenhuizen in Ref. \cite{VanNieuwenhuizen:1973fi}, employed in Ref. \cite{Nunez:2004ji} to study the class of $f(R)$ models and recently generalized in Ref. \cite{Biswas:2011ar} for general metric theories of gravitation involving arbitrary combinations of curvature invariants at an arbitrary order in derivatives. The main result is that the propagator $\Pi$ for metric fluctuations around flat background can be written, in the Fourier space, as
\be
k^2\Pi =  \frac{\mathcal{P}^2}{a(-k^2)} - \frac{\mathcal{P}^0}{ a(-k^2)-3c(-k^2)}\,,
\ee
where $\mathcal{P}^2$ picks up the spin-2 and $\mathcal{P}^0$ the scalar modes of the fluctuations.

Let us start by considering the class of $f(R,\hat{R})$ theories. Without significant loss of generality, we may assume that $f^{(0)}_{,R}+f^{(0)}_{,\hat{R}}=1$, so that we recover Einstein's General relativity for the graviton propagator at the infrared limit. The sum may converge to another positive constant than unity in this limit, but taking that into account amounts to an overall rescaling which is irrelevant for our purposes.
We readily see that then
\be
a=-b=1\,, \quad c=-d=1-2(\F +4\f )\Box+6A\f\Box^2\,, \quad e=2(\F+4\f )\Box-6A\f\Box^2\,.
\ee  
The propagator we then obtain is 
\be \label{frrp}
\Pi_{f(R,\hat{R})}=\Pi_{GR} +\frac{3\lp \F + 4\f +3A\f k^2\rp}{2\lp 1+3\F k^2 + 12 \f k^2 +9 A\f k^4\rp} \mathcal{P}^0\,,
\ee
where the GR propagator is given by
\be
\Pi_{GR} = \frac{1}{k^2}\lp \mathcal{P}^2-\frac{1}{2}\mathcal{P}^0\rp\,.
\ee
A subtlety here is that though the scalar propagator comes with the wrong sign, it does not imply a ghost but rather cancels the unphysical longitudinal degree of freedom contained in the spin-2 propagator $\mathcal{P}^2$ \cite{VanNieuwenhuizen:1973fi, Biswas:2013ds}.
Let us first check that the result agrees with the previously known results in other appropriate limits and then proceed to study the new generalizations.

\subsection{Metric $f(R)$ models}

In the pure metric $f(R)$ case, $\f=A=0$ and we have \cite{Nunez:2004ji}
\be \label{scalaron}
\Pi_{f(R)}=\Pi_{GR} +\frac{1}{2\lp k^2+ (3\F)^{-1}\rp}\mathcal{P}^0\,.
\ee
Thus we have an extra scalar degree of freedom, as we expect since the $f(R)$ models are known to be equivalent to Brans-Dicke theories with a vanishing parameter $\omega_{BD}=0$. 
The mass of the ''scalaron'' is $m^2= (3\F)^{-1}$, and as long as $f''(R)>0$ the theory is stable, otherwise a tachyonic mass spoils the stability around Minkowski space. This agrees with the stability condition found in \cite{Faraoni:2006sy} and generalized in \cite{Amendola:2010bk}. 

\subsection{Palatini $f(\hat{R})$ models}

It is well known that the Palatini-type $f(\hat{R})$ models are equivalent to Brans-Dicke theories with the parameter $\omega_{BD}=-3/2$. This particular value corresponds to vanishing kinetic term of the field, which thus is nondynamical.  Therefore we expect that no additional scalar degree of freedom should appear. As previously, we may assume that $f^{(0)}_{,\hat{R}}=1$, and we have now of course that $\F=\f=f^{(0)}_{,\hat Q}=0$. Hence,
\be
\Pi_{f(\hat{R})}=\Pi_{GR}\,,
\label{013}
\ee
confirming our expectation.

A comment here has to be made for the sake of generality. Recall that we assumed the condition $f^{(0)}_{\hat{R}\hat{R}}=0$ which actually restrict our analysis in the nonlinear Palatini case. However it easy to realize that Palatini $f(R)$ gravity adds no degrees of freedom from some considerations in absence of matter fields. In vacuum the $f(\hat R)$ gravity field equations are
\begin{align}
f_{,\hat{R}}\hat{R}_{\mu\nu} -\frac{1}{2}g_{\mu\nu}f = 0 \label{011} \,,\\
\hat\nabla_\lambda\left(\sqrt{-g}f_{,\hat{R}}g^{\mu\nu}\right)=0 \label{012} \,.
\end{align}
Once the function $f$ is prescribed, the trace of (\ref{011}) will give an algebraic equations for $\hat R$ whose solutions will generally be $\hat R=\hat R_0=$ const. This implies that $f_{,\hat R}$ will be a constant too and then equation (\ref{012}) will reduce the independent connection to Levi-Civita. The field equations (\ref{011}) will then become the Einstein field equations with a cosmological constant in vacuum which is a well known result of Palatini $f(R)$ gravity. The propagator of the theory will thus be equal to the GR one, as shown in (\ref{013}), even if the $f^{(0)}_{\hat{R}\hat{R}}=0$ constraint is relaxed.

\subsection{Hybrid $f(X)$ models}

The lagrangian of the form $R+f(\hat{R})$ results in field equations where the correction to Einstein gravity are controlled by the failure of the GR trace equation $X=R-\kappa^2 T$. They belong to the algebraic class of scalar-tensor theories which single out the unique interpolation of the metric and Palatini $f(R)$ theories by requiring that the scalar field has an algebraic expression in terms of the trace and the Ricci scalar, $\phi=\phi(T,R)$ \cite{Koivisto:2009jn}. It was mentioned in Ref. \cite{Amendola:2010bk} that this produces the peculiar $f(X)$ type of field equations, and their implications and phenomenology has been recently studied in contexts of Solar system constraints, cosmology and wormholes \cite{fxliterature}. It was already remarked in \cite{Koivisto:2009jn} that in Ricci-flat spacetimes the $f(X)$ theories share the properties of Palatini $f(R)$ theories, which in vacuum reduce to GR with a possible cosmological constant. 
Therefore it is not a surprise that we find no new propagating degrees of freedom in Minkowski vacuum,
\be
\Pi_{f(X)}=\Pi_{GR}\,.
\label{016}
\ee
 Interestingly though, this class of theories is not equivalent to either of the previous two cases, since when one considers curved spacetimes, a new scalar degree of freedom appears. In this sense, the $f(X)$ gravity is a quite minimalistic scalar-tensor extension of GR, as the scalar propagates only in the presence of curvature. For the viability criteria for these models in regards of cosmology and Solar system, we refer the reader to \cite{fxliterature}.

Again we notice that the constraint $f^{(0)}_{\hat{R}\hat{R}}=0$ prevents a complete analysis also in this case. However we can repeat the same argument above and consider the $f(X)$ field equations in vacuum
\begin{align}
R_{\mu\nu}-\frac{1}{2}g_{\mu\nu}R +f_{,\hat{R}}\hat{R}_{\mu\nu} -\frac{1}{2}g_{\mu\nu}f = 0 \label{014} \,,\\
\hat\nabla_\lambda\left(\sqrt{-g}f_{,\hat{R}}g^{\mu\nu}\right)=0 \label{015} \,,
\end{align}
where now the Lagrangian is $R+f(\hat R)$ and the function $f$ only depends on the Palatini curvature scalar $\hat R$. If we now take the trace of (\ref{014}) we obtain, once the function $f$ is prescribed, an algebraic equation relating $\hat R$ to $R$. The function $f_{,\hat R}$ will generally become a function of $R$ and equation (\ref{015}) will reduce to its (metric) $f(R)$ correspondent. The propagator analysis will then equal the metric $f(R)$ gravity one meaning that in the end we will find a scalar propagating degree of freedom in addition to the graviton. As noted in \cite{fxliterature} the theory is thus quite different from Palatini $f(R)$ gravity presenting a richer phenomenology. Note that when the metric Ricci scalar vanishes, as in Minkowski spacetime, the trace of equation (\ref{014}) will again give the solution $\hat R=\hat R_0$ and the scalar degree of freedom will disappear in agreement with (\ref{016}). As said above the $f(X)$ gravity phenomenology differs from the $f(\hat R)$ gravity only in curved spacetimes.

\subsection{The hybrid $f(R,\hat{R})$ models}

The generalized hybrid Ricci scalar theories were introduced in \cite{Tamanini:2013ltp,Flanagan:2003iw} and found to have qualitatively different properties compared to the more restricted class of $f(X)$ models described above. In particular, the $f(R,\hat{R})$ were shown to be equivalent to a class of biscalar-tensor theories. It is indeed seen from our formula (\ref{frrp}) that these theories have an extra $\mathcal{P}^0$ spin-0 propagator with a double pole, corresponding to two propagating scalar degrees of freedom. From the formula, we can easily deduce the masses of these scalar fields. We get
\be
%m^2_a  & = &  \frac{\F+4\f+S }{6A\f}\,, \\
%m^2_b  & = &  \frac{2}{3\lp \F+4\f+S\rp}\,,
m^2_\pm = \frac{f^{(0)}_{,\hat{R}}}{18\lp \f\rp^2}\lp \F+4\f \pm S\rp\,,
\ee
where we have defined for convenience
\be
S  \quad \equiv \sqrt{\lp\F+4\f\rp^2- 12\frac{\lp\f\rp^2}{f^{(0)}_{,\hat{R}}} }\,.
\ee
We note that the scalar particle with mass squared $m_-^2$ corresponds to the scalaron appearing in (\ref{scalaron}) in the limit of pure $f(R)$ gravity, but in general now has a shifted mass. The other scalar is a new particle that occurs due to nontrivial dependence upon $\hat{R}$, and unlike in the case of $f(X)$ gravity, it propagates also in Ricci-flat spaces. The condition that neither of the scalars has a tachyonic instability, is given by
\be \label{frrc}
f^{(0)}_{,\hat{R}}  >  0\,, \quad \mbox{and} \quad
%\F+4\f+\sqrt{\lp\F+4\f\rp^2-\frac{4\lp\f\rp^2}{3f^{(0)}_{,\hat{R}}}}> 0\,.
\F+4\f-S> 0\,.
\ee
The residues at the two poles corresponding to these masses are
\ba
%r_a & = & \frac{-\F-4\f + S}{4S}\,, \\ 
%r_b & = & \frac{\lp\F+4\f\rp \lp \F+4\f + S \rp - \frac{2\lp\f\rp^2}{3f^{(0)}_{,\hat{R}}} }{2\lp  \lp\F+4\f\rp \lp \F+4\f + S \rp - \frac{4\lp\f\rp^2}{3f^{(0)}_{,\hat{R}}} \rp}\,.
r_\pm = \frac{S \pm \lp\F+ 4\f \rp}{4S}\,.
\ea
In order for neither of these scalars to be a ghost, we should have both $r_+>0$ and $r_->0$. The second condition would require that
\be
\F+4\f-S < 0\,,
\ee
in contradiction with (\ref{frrc}). It seems then that we cannot avoid both tachyons and ghosts in this theory.

Again we recall that we have assumed $f^{(0)}_{\hat R\hat R}=0$ in our analysis. It might be that relaxing this constraint allows for viable $f(R,\hat R)$ theories. This would anyway going to depend on the matter degrees of freedom since $\hat R$ can only be replaced in the field equations solving the trace of (\ref{003}). Besides the problems we already mentioned which disappear with the $f^{(0)}_{\hat R\hat R}=0$ ansatz, a complete analysis for this class of theories cannot be performed without explicitly specifying the function $f$ and a matter action.
%For example, if $f^{(0)}_{,\hat{R}}>0$, either the field $b$ is a tachyon or the field $a$ is a ghost. 

\subsection{The hybrid Ricci-squared $f(\hat{R},\hat{Q})$ theories}

Let us finally consider the $\hat{Q}_H$-invariant. For simplicity, we restrict to models here without nonlinear dependence on the metric Ricci scalar; it is easy to see that this does not affect our conclusions essentially. Basically the graviton propagator acquires its structure from the function $a(\Box)$ in (\ref{abox}), and now only the higher-derivative term $\hat{Q}_H$ modifies it. We can arrange the result for the propagator in the form 
\begin{align}
\Pi_{f(\hat{R},\hat{Q})} = \frac{\Pi_{GR}}{\lp 1-\frac{1}{4}\lp\Q\rp^2 k^4\rp} +\frac{3\Q\lp 1+\frac{3}{4}\Q k^2\rp}{2\lp 1-\frac{1}{4}\lp\Q\rp^2k^4\rp\lp 1+3\Q k^2+2\lp\Q\rp^2 k^4\rp}\mathcal{P}^0 \,.
\end{align}
The sixth order theory we have at hand has a modulated graviton propagator which adds two extra poles. In addition, there appears a scalar propagator that has five poles. This is in a quite drastic contrast with respect to the metric $Q$-theory which contains only one additional spin-2 particle and features fourth order field equations. We need not analyze in detail the properties of the new degrees of freedom here, since it is obvious the theory as such is seriously haunted by ghosts and thus not physical. 

\section{Conclusions}
\label{sec:concl}

We considered theories of gravity
%in the first order, i.e. the Palatini formulation that lets the connection to vary independently of the metric, 
in the generalized hybrid framework where, besides the independent Palatini connection, the metric Levi-Civita connection is also allowed to enter into the fundamental action. In particular, we applied the formalism of \cite{Biswas:2011ar, Biswas:2013ds, VanNieuwenhuizen:1973fi,Nunez:2004ji} to uncover the classical and quantum stability of the theories, i.e.~to find out which of the theories may be free of ghosts and tachyonic instabilities. 

To this end, we started with an action involving three types of curvature invariants constructed from the metric and the independent connection, and brought the general theory into a form of an effectively purely metric theory at the relevant limit. The generalisation of the analysis to other types of invariants is not entirely straightforward because an algebraic solution for the independent connection in terms of the metric and the curvature invariants may not be generically available, but instead one would need to supplement the system with differential equations of motion for the connection.

For the simplest theories we reproduced the well known results: the metric $f(R)$ is equivalent to a scalar-tensor theory and stable given $f''(R)>0$, whilst the Palatini-type $f(\hat{R})$ theory carries no extra degrees of freedom. Theories involving both Ricci invariants show quite interesting properties: the $f(X)$ theories with only linear dependence upon the metric $R$, are viable and in fact reduce to GR in flat spacetime, whereas more general theories with nontrivial dependence upon both of the Ricci invariants inevitably contain instabilities. They introduce two new scalar degrees of freedom, where either one of them is a ghost or the other has a negative mass-squared associated to its excitations around a flat background. 

It is well understood that in the metric formalism, the $f(R)$ theories are a special class of viable GR extensions  \cite{Woodard:2006nt} - what we have found here is that the $f(X)$ gravity seems to have a similar unique place amongst the space of all theories within the hybrid formalism. Though we have, for the technical reason stated above, explicitly derived the propagator only for theories involving the $\hat{Q}$ besides the Ricci scalar invariants, from the general structure of the equations we expect that similar conclusions would hold quite generically for all the hybrid theories. Discovering either exceptions (as the Gauss-Bonnet term in the metric formulation) or a robust no-go theorem, is a challenge for future studies.  

\acknowledgments

TK is supported by the Research Council of Norway.

\end{document}